# Coherent control of nitrogen-vacancy center spins in silicon carbide at room temperature

Jun-Feng Wang[1,2], Fei-Fei Yan[1,2], Qiang Li[1,2], Zheng-Hao Liu[1,2], He Liu[1,2], Guo-Ping Guo[1,2], Li-Ping Guo[3], Xiong Zhou[3], Jin-Ming Cui[1,2], Jian Wang[1,2], Zong-Quan Zhou[1,2], Xiao-Ye Xu[1,2], Jin-Shi Xu[1,2,*], Chuan-Feng Li[1,2,*] and Guang-Can Guo[1,2]

*[1]CAS Key Laboratory of Quantum Information, University of Science and Technology of China, Hefei, Anhui 230026, People's Republic of China*

*[2] CAS Center for Excellence in Quantum Information and Quantum Physics, University of Science and Technology of China, Hefei, Anhui 230026, People's Republic of China*

*[3]Accelerator Laboratory, School of Physics and Technology, Wuhan University, Wuhan, Hubei 430072, People's Republic of China*

[*]Corresponding author: jsxu@ustc.edu.cn, cfli@ustc.edu.cn

**Abstract**

**Solid-state color centers with manipulatable spin qubits and telecom-ranged fluorescence are ideal platforms for quantum communications and distributed quantum computations. In this work, we coherently control the nitrogen-vacancy (NV) center spins in silicon carbide at room temperature, in which telecom-wavelength emission is detected. We increase the NV concentration six-fold through optimization of implantation conditions. Hence, coherent control of NV center spins is achieved at room temperature and the coherence time $T_2$ can be reached to around 17.1 μs. Furthermore, investigation of fluorescence properties of single NV centers shows that they are room temperature photostable single photon sources at telecom range. Taking advantages of technologically mature materials, the experiment demonstrates that the NV centers in silicon carbide are promising platforms for large-scale integrated quantum photonics and long-distance quantum networks.**

Identification and coherent control of novel spin defects are important for extending the scope of solid-state quantum information science [1-3]. Optically active defect spins in solid–state systems have been widely used in quantum photonics, quantum communications, quantum computation and quantum metrology [1-22]. Nitrogen-vacancy (NV) centers in diamond have become leading candidates due to their excellent properties, including photostability and long spin coherence times even at room temperature [4,5]. However, the drawbacks of visible-wavelength emission and lack of a mature nanofabrication method for diamond limit their application to long-distance quantum communications and wafer-scale quantum technology [1-6]. To overcome these drawbacks, in recent years, defects in silicon carbide (SiC) have been developed as promising platforms for the quantum information science [6-22].

SiC has been widely used in power electronic devices and has commercially available inch-scale growth and matured nano-fabrication protocols [6-13]. Several bright (about Mcps) visible and telecom range single photon emitters have been found in different polytypes of SiC, which can be used for quantum photonics and quantum communications [7-10]. Moreover, as with NV centers in diamond, there are also optically active spin defects: silicon vacancy and divacancy defects in SiC, which can be polarized by laser and controlled by microwave [6,11-22]. These two kinds of defects have realized the single spin manipulation with long coherence times (about 1 ms) [11-16], high-fidelity near infrared spin-to-photon interface [16,17] and high-sensitivity quantum metrologies for magnetic fields [18], electric fields [19], local strain fields [20] and temperature [21,22] etc. However, their emission spectra are only in the near infrared [6,11-16]. Efficient generation and coherent control of optically active spin defects with telecom-range emissions in SiC are still great challenges.

Most recently, NV centers in 4H-SiC and 3C-SiC have been demonstrated to be electron paramagnetic defects with emission at telecom wavelengths [23-30]. The $N_C V_{Si}$ center in SiC consists of a nitrogen impurity substituting a carbon atom ($N_C$) and a silicon vacancy ($V_{Si}$) adjacent to it [23-30]. However, most previous experiments have focused on the properties of low temperature (LT) photoluminescence (PL) spectra and electron paramagnetic resonance (EPR) of the NV centers [23-30]. Little is known about the optically-detected-magnetic-resonance (ODMR) spectrum and spin coherence property. Moreover, there have yet to be report on the scalable generation of single NV centers in SiC, which is vital for constructing on-chip quantum processors [6,11-17].

In this work, we realize coherent control of the NV center spins at room temperature and scalable generation of the single NV centers in 4H-SiC. By optimizing the implanted conditions, the concentration of NV centers ensemble increases by about six times. The PL spectra of NV centers show that the wavelengths are in the telecom range. We then implement the ODMR measurement and realize coherent control of the NV center ensemble spins at room temperature with a coherence time $T_2$ of around 17.1 μs. Moreover, we find that the dephasing time $T_2^*$ decreases as the nitrogen implanted dose increases from $1\times10^{13}$ cm$^{-2}$ to $1\times10^{16}$ cm$^{-2}$. Finally, we present the implanted single NV center and characterize the fluorescence property. Our experiments pave the way for using the NV centers in SiC for quantum photonics and quantum information processing.

In the experiment, a bulk high-purity 4H-SiC epitaxy layer sample is used [31,32]. To generate the shallow NV centers (60 nm, SRIM) in 4H-SiC, 30 keV nitrogen ions are implanted. The experiments are performed in a homebuilt confocal setup. A 1030 nm laser is used to excite the NV centers since this pumping wavelength has a better exciting effect for NV centers and reduces the emission from the divacancy defects at the same time [24,27]. For the room temperature confocal setup, an oil objective with 1.3 N.A. is used to excite the NV centers. The fluorescence is collected by a multimode fiber to a photoreceiver (Femto, OE-200-Si) after a 1150 nm long pass (LP) filter for the spin control experiment of the NV center ensemble [21]. The spin signal is detected by the lock-in methods [6,21,32]. To investigation of the single NV defect, the fluorescence is collected by a single-mode fiber to a superconducting single-photon detector (Scontel) [22]. A Montana cryostation combined with a confocal setup is used for the low temperature experiment [21,32].

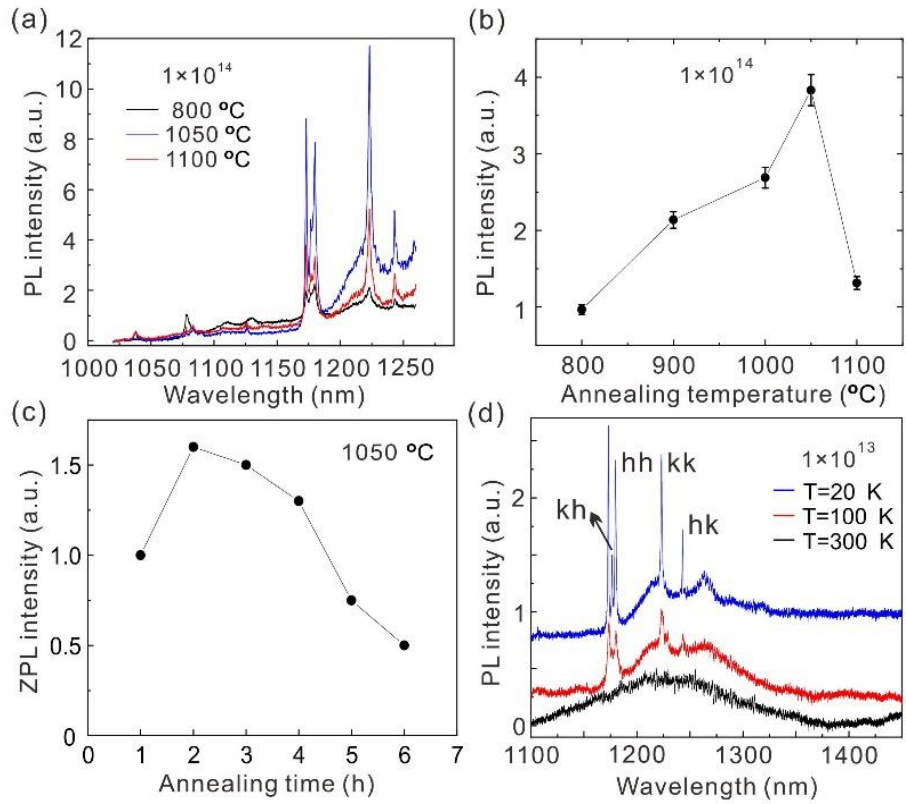

FIG. 1. (a) The LT-PL spectra of the NV centers and divacancy defects as a function of the annealing temperature. (b) The mean counts of the scanning images (10×10 μm$^2$) as a function of the annealing temperature with a pump laser power of 1 mW. (c) The NV center ZPL intensity as a function of the annealing time at 1050 °C. (d) The PL spectra of NV centers as the sample temperature increases from 20 K to 300 K.

Since the divacancies are intrinsic vacancy-related defects, it is inevitable that a few will be generated during the nitrogen implantation process [24,27]. Moreover, the PL and ODMR spectra of the NV centers in 4H-SiC are very close to that of the divacancy defects. In order to efficient fabricate single NV centers, it is critical to optimize the implanted conditions to efficiently generate the NV centers. In the experiment, the samples implanted with nitrogen ions with an influence of $1\times10^{14}$ cm$^{-2}$ are annealed at different temperature for the same time (1 hour). To measure the whole PL spectra of the sample, a 980 nm laser was used in the PL experiments. Figure 1a shows the low temperature (20K) PL spectra of the implanted samples at different annealing temperature with a 1000 nm LP filter. The results show that the PL intensities of the zero phonon lines (ZPLs) of NV centers obviously increase as the annealing temperature increases from 800 °C to 1050 °C, while the PL intensities of the divacancy defects decrease at the same time, as shown in the Supplemental Material (Fig. S1b) [33]. The corresponding scanning images of 10×10 μm² areas using the 1150 nm LP filter with a laser power of 1 mW are also investigated [33]. The mean counts of the scanning images (10×10 μm²) are presented in Fig. 1b, these counts increase by around four times as annealing temperature increases from 800 °C to 1050 °C; then they decrease to about 1.3 times their original value at 1100 °C, which is similar to the previous deep depth NV center implantation results [30]. Furthermore, as presented in Fig. 1c, the optimal ZPL intensity for an annealing time of 2 hours is 1.6 times larger than that for 1 hour. Thus, the NV center concentration increase six-fold through optimization of the implantation conditions.

In addition, we also study the PL spectra of the sample ($1\times10^{13}$ cm$^{-2}$) as a function of the sample temperature increasing from 20 K to 300 K (Fig. 1d). We can see that the ZPL peaks of the NV centers decrease as the sample temperature increases and they can still be observed when the temperature increases to 150 K [33]. At room temperature, the PL spectra range from 1100 nm to about 1420 nm, which covers the O-band and E-band telecom range for around 45 % percent.

The electronic ground state of the negative nitrogen-vacancy (NV$^-$) center in 4H-SiC is a spin-1 state [23-29]. There are two types of bond directions for NV centers. One is c-axis defects (hh and kk, $C_{3v}$ symmetry), and another is basal defects (hk and kh, $C_{1h}$ symmetry) [28]. The spin Hamiltonian is:

$$H = D[S_z^2 - S(S+1)/3] + E(S_x^2 - S_y^2) + g\mu_B B S_z, \qquad (1)$$

where *D* and *E* are the axially symmetric and anisotropic components of the zero-field-splitting (ZFS) parameter, respectively, *g = 2* is the electron g-factor, $\mu_B$ is the Bohr magneton and *B* is the applied axial static magnetic field. Moreover, they also have $^{14}$N nuclear (I = 1) hyperfine interactions, which is around 1.2 MHz.

In the experiment, we only focus on the NV center (hh). First, ODMR measurement of the NV center (hh) ensemble ($1\times10^{14}$ cm$^{-2}$ sample) with low laser and microwave power are performed at 20 K (bottom part, Fig. 2a) and room temperature 300 K (upper part, Fig. 2a), respectively. The obvious three splitting with at 1.3 MHz under both temperatures is due to the nitrogen nuclear hyperfine coupling. The low temperature ZFS value is 1330.4 ± 0.1 MHz and the hyperfine coupling A is around 1.3 MHz, which is same with previous EPR reports of the NV center (hh) [28]. In addition, the room temperature ZFS is around 1319.0 ± 0.1 MHz, which is 11.4 MHz smaller than the LT value. To further identify the bond direction of the observed defects spins. The magnetic

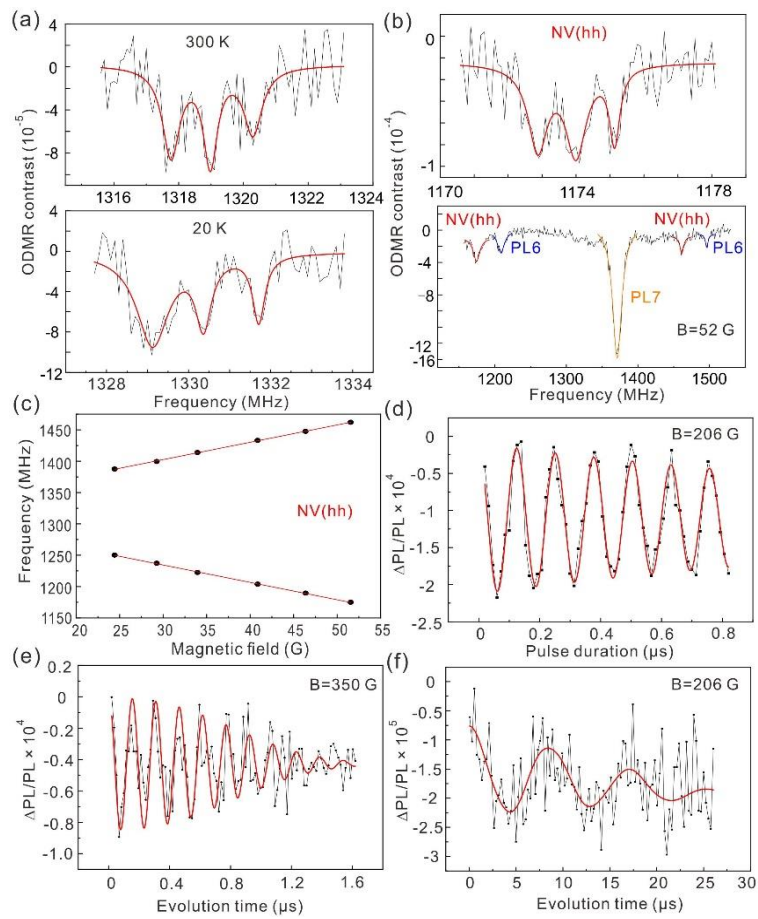

FIG. 2. (a) The ODMR measurement of the NV center (hh) ensemble at low (bottom part, 20K), room (upper part, 300K) temperature and zero magnetic field. The red line are the fitting using the Lorentzian function. (b) The bottom part shows the ODMR measurement of the NV center ensemble at a magnetic field of 52 G. And PL 6 and PL7 are divacancy defects in 4H-SiC. The three peaks in the upper part correspond to the nitrogen hyperfine. (c) The ODMR resonant frequencies of the NV

center ensemble as a function of the axial magnetic field. (d) The spin Rabi oscillation between $|0\rangle$ and $|-1\rangle$ states for the c-axis NV centers (hh) at room temperature. (e) The free induction decay of the NV center spins at room temperature. The red line is the fits of the data using the decayed sinuous function [33]. (f) Hahn echo measurement of the NV center ensemble.

field dependent ODMR measurement is also performed. The bottom part of Fig. 2b shows the ODMR measurement at 52 Gauss and the upper part shows the corresponding nitrogen hyperfine coupling. PL6 and PL7 are divacancy defects in 4H-SiC [11]. The summary of the two branch resonant frequencies as a function of the magnetic field is shown in Fig. 2c. The red lines are the linear fitting to the data with slopes of 2.8 MHz/G. The corresponding ZFS values is around 1318.5 MHz, which is consistent with the directly measured value. These results demonstrate that the investigated spin is the c-axis NV center (hh). We further perform the angle-resolved hyperfine measurements by changing the angle of the external magnetic field. The results are displayed in the supplementary materials [33] and are consistent with the results shown here.

Coherent control of the spin state at room temperature is the cornerstone of NV center applications in quantum information processing [6,11-14]. Thus, we measure the Rabi oscillation of the NV center ensemble using a resonant microwave frequency of 742.3 MHz between $|0\rangle$ and $|-1\rangle$ states transition at 206 Gauss using the standard pulse sequences [6, 21, 33]. The corresponding coherent Rabi oscillation of the NV centers at room temperature is shown in Fig. 2d. Inferred from the fit, the Rabi frequency is about 7.9 MHz. The obvious oscillation signals demonstrate the coherent control of the NV center spins at room temperature [6,11,12,14]. The spin coherence properties of the NV centers in SiC are important for quantum computation and high sensitivity quantum sensing [6,11,15,18-22,35,36]. Since increase the magnetic field could make the Ramsey fringe appear more clearly [37], we measure the free induction decay of the NV centers at room temperature with a magnetic field of around 350 G. The Ramsey fringe is shown in Fig. 2e. The inferred dephasing time $T_2^*$ is 1.0 ± 0.1 μs [33], which is comparable with the silicon vacancy [38] and divacancy [14,16] in SiC. As presented in Fig. 2f, the Hahn echo is also investigated and the periodic modulations of its envelope are due to the $^{29}$Si and $^{13}$C nuclear spin bath. The inferred coherence time $T_2$ = 17.1 ± 4.0 μs [33], which is comparable with the previous room temperature divacancy defects ensemble and silicon vacancy ensemble results [6,39].

We further investigate the influence of the implanted doses on the coherence properties [11,35]. The LT-PL spectra of NV centers for different doses are shown in Fig. 3a, which both have obvious NV center ZPLs. Figure 3b shows mean counts of the NV centers as a function of the dose. The counts increase with the dose from $1\times10^{12}$ cm$^{-2}$ to $1\times10^{15}$ cm$^{-2}$, while slightly decreasing when the dose increases to $1\times10^{16}$ cm$^{-2}$. The saturation of the counts may be due to the nitrogen ion implanted damage to crystal lattices, leading to amorphization of the SiC [32]. We also study the changes of the width of NV centers (hh) ZPL as the dose increase. As shown in Fig. 3c, the ZPL width slowly increases from 1.2 nm to 1.3 nm as the dose increases from $1\times10^{13}$ cm$^{-2}$ to $1\times$ $10^{14}$ cm$^{-2}$, before rapidly increasing to 3.0 nm as the dose increases to $1\times10^{16}$ cm$^{-2}$. Moreover, we compare the free induction decay of the sample with the different implanted dose. The NV center ensemble dephasing time $T_2^*$ as a function of implanted dose is presented in Fig. 3d. The $T_2^*$ decreases slightly as the dose increases from $1\times10^{13}$ cm$^{-2}$ to $1\times10^{15}$ cm$^{-2}$, then it quickly decreases to 0.45 μs as the dose increases to $1\times10^{16}$ cm$^{-2}$. The decrease of the $T_2^*$ may be due to the increase of the defects densities and the damage to lattice of the SiC [11].

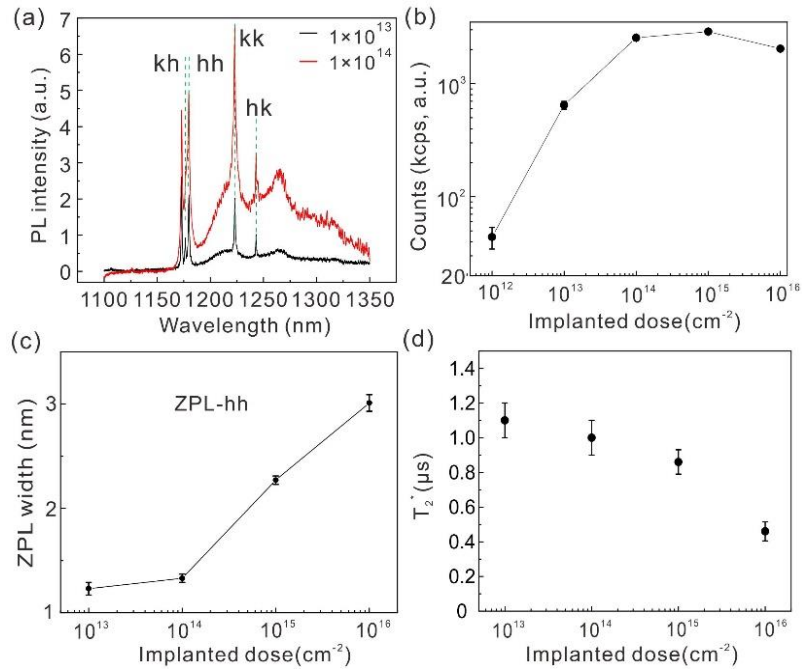

FIG. 3. (a) The LT (20 K) PL spectra of the NV centers ensemble with different implanted doses. (b) The mean counts of the NV centers on samples with different implanted doses. (c) The widths of ZPL of the NV center (hh) as a function of the implanted dose. (d) Comparison of the dephasing time, $T_2^*$ for samples with four different implanted doses.

Generating single NV centers in SiC is vital for various quantum technologies, including quantum photonics, quantum networks and nanoscale quantum sensing [5,12-14,16,17,31,32,40,41]. In our experiment, a 200-nm-thick polymethyl methacrylate (PMMA) layer is deposited on the SiC surface. Then, 70 ± 10 nm diameter nano-aperture arrays (2 × 2 μm$^2$) are generated using electron-beam lithography technology [31,32]. 30 keV nitrogen ions with dose of $2.5\times10^{11}$ cm$^{-2}$ are implanted through the nano-apertures to generate a single NV center array in 4H-SiC.

We then characterize the fluorescence properties of the generated single NV centers. A representative 20 × 20 μm$^2$ confocal image of the single defect arrays is displayed in Fig. 4a. To identify the number of NV center, we perform the Hanbury-Brown and Twiss (HBT) measurement and obtain the corresponding second-order photon correlation function. Figure 4b presents the result for the circled defects in Fig. 4a. Then we fit $g^2(\tau)$ using the function $g^2(\tau) = 1-a*exp(-|\tau|/\tau_1)+b*exp(-|\tau|/\tau_2)$, where *a* and *b* are fitting parameters and $\tau_1$ and $\tau_2$ are related to the excited and metastable state lifetimes, respectively. The obtained $g^2(0)$ is around 0.25, which demonstrates that it is a single defect [33]. Inferred from the fit, the $\tau_1$ and $\tau_2$ are about 2.4 ns and 450 ns, respectively. Moreover, we measure the fluorescence lifetime of the NV centers in the same sample using a 1040 nm femtosecond pulse laser at both room temperature and low temperature. Figure 4c shows the lifetime of the NV center ensemble at room temperature. The red line is the fit of the data using a double-exponential decay function. The fluorescence lifetime is about 2.1 ± 0.1 ns which is almost the same with the short excited state lifetimes $\tau_1$ [42,43]. The lifetime is much smaller than that of the divacancy in 4H-SiC

[14,16,20,44].

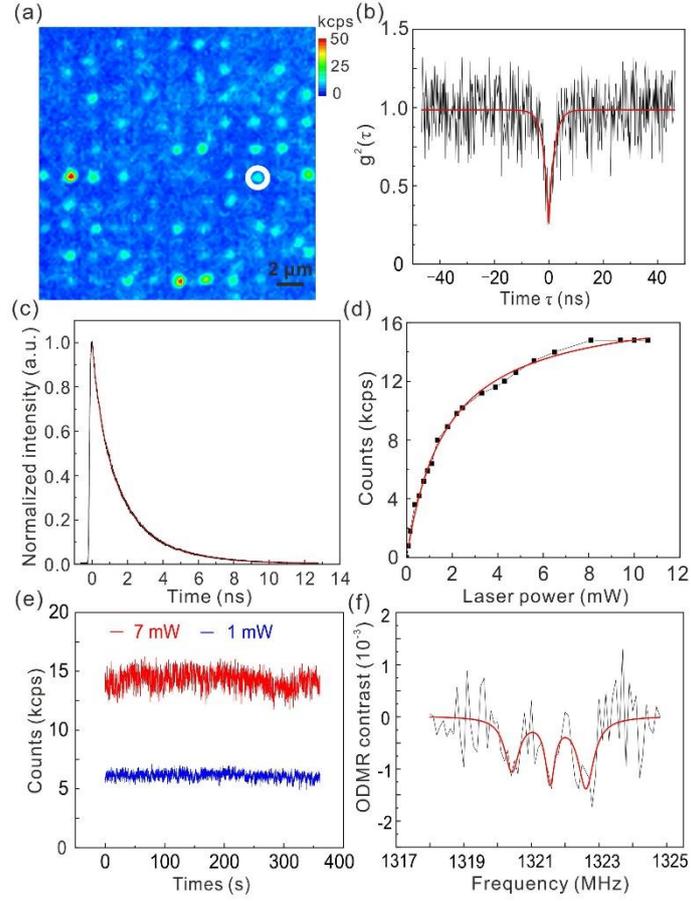

FIG. 4. (a) A confocal scan image (20×20 μm$^2$) of the single NV center arrays in 4H-SiC with a laser power of 4 mW. The circled point is the investigated single NV center. The scale bar is 2 μm. (b) The correlation function measurement of the circled single NV center in (a). (c) The lifetime of the NV centers ensemble at room temperature. (d) The counts of the single NV centers as a function of laser power. (e) The photostability of the single NV center at two different pump laser power with the time bin set to 100 ms. (f) The ODMR measurement of a single NV center.

The saturation curve of a single NV center is shown in Fig. 4d. The red line is the fitting using the power dependence model $I(P) = I_s/(1 + P_0/P)$, where $I_s$ is the maximal emission counts and $P_0$ is the saturation power. The fit indicates that $I_s$ is around 17.4 ± 0.2 kcps and the saturation power $P_0$ is around 1.7 ± 0.1 mW, respectively. The saturation count is comparable with silicon vacancy and divacancy in 4H-SiC [12,16]. The low count rate might come from the trapping in other dark states [31]. Figure 4e exhibits time trace of counts of the single NV centers for about 6 minutes with a 100 ms time bin for two different pump laser powers. The stable fluorescence emission demonstrates that the NV center is a stable infrared-wavelength single photon source at room temperature. Furthermore, we also measured the single NV center ODMR (Fig. 4f). Inferred from the fitting, the ZFS is 1321.5 MHz and the little difference from the ensemble value may due to the different residual strains in different samples [31]. In addition, the nitrogen hyperfine coupling is around 1.1 MHz, which is consistent with previous EPR reports [28].

In conclusion, we coherently control the NV center spins and characterize the fluorescence and ODMR properties of a single NV centers in 4H-SiC at room temperature. Through optimizing the annealing condition, the concentration of the NV centers increases about six times. Furthermore, we coherently control the spin states of the NV centers (hh) ensemble at room temperature and obtain a coherence time $T_2$ of around 17.1 μs, which is comparable with that of the divacancy and silicon vacancy defects in SiC [6,39]. The dephasing time $T_2^*$ is shown to decrease as the implanted dose increases. The single NV center are demonstrated to be room temperature photostable with a saturation count of around 17.4 kcps, which is comparable with divacancy [14,16] and silicon vacancy [12,13,31,32] in 4H-SiC. The experiment paves the way for using NV centers in technologically friendly SiC materials in quantum photonics, quantum information processing and scalable quantum networks.

**Acknowledge**

We thank Prof. J. Wrachtrup for the helpful discussion. We also thank Yongxiang Zheng, Jun Hu for their help in the experiment. This work was supported by the National Key Research and Development Program of China (Grant No. 2016YFA0302700 and 2017YFA0304100), the National Natural Science Foundation of China (Grants No. 61725504, 61905233, 11975221, 11804330, 11821404, 11774335 and U19A2075), the Key Research Program of Frontier Sciences, Chinese Academy of Sciences (CAS) (Grant No. QYZDY-SSW-SLH003), Science Foundation of the CAS (No. ZDRW-XH-

2019-1), Anhui Initiative in Quantum Information Technologies (AHY060300 and AHY020100), the Fundamental Research Funds for the Central Universities (Grant No. WK2030380017 and WK2470000026).